# An outlook on the estimate of the Solar Quadrupole Moment from Relativistic Gravitation Contributions

J.P. Rozelot,[1] A. Kilcik[2] and Z. Fazel[3]

[1] *Université de la Côte d'Azur, CNRS-OCA, 77, Chemin des Basses Moulières, 06130 Grasse, France.*
[2] *Akdeniz University, Faculty of Science, Dpt of Space Science & Technologies, 07058, Antalya (T).*
[3] *Faculty of Physics, University of Tabriz, Tabriz, Iran.*

**ABSTRACT**
Of all the solar fundamental parameters (mass, diameter, gravity at the surface, angular momentum...), the gravitational moments have been quite often ignored in the past, mainly due to the great difficulty to get a reliable estimate. Even though the order of magnitude of the solar quadrupole moment $J_2$ is now known to be $\approx 2.10^{-7}$, its accurate value is still discussed. Furthermore, stellar equations combined with a differential rotation model, as well as the inversion techniques applied to helioseismology, are methods which are solar model dependent, *i.e.* implying solar density and rotation laws. Hence the need for checking new ways in estimating the dynamical solar quadrupole moments based on the motion of spacecrafts, celestial bodies or light in the gravitational field of the Sun. Indeed, the expansion in multipoles $J_{(l,\ l=2,...)}$ of the gravitational potential of a rotating body affects the orbital motion of planets at a relativistic level. We will recall here the recent progresses made in testing General Relativity through the contribution of the first solar quadrupole moment. Using the Eddington-Robertson parameters, we recall the constraints both on a theoretical and experimental point of view. Together with $\gamma$, which encodes the amount of curvature of space-time per unit rest-mass, the Post–Newtonian Parameter $\beta$ contributes to the relativistic precession of planets. The latter parameter encodes the amount of non–linearity in the superposition law of gravitation. Even though in principle, it would be possible to extract $J_2$ from planetary ephemerides, we observe that it is significantly correlated with other solution parameters (semi-major axis of planets, mass of asteroids...). Focusing on the $J_2$ correlations, we show that in general, when $\beta$ and $\gamma$ are freed, the correlations [ $\beta$, $J_2$] and [$\gamma$, $J_2$] are $\approx$ 45% and $\approx$ 55% respectively. The situation could be improved with additional spacecraft measurements but remains difficult. Moreover, all the planetary dynamics–based values are biased by the Lense–Thirring effect, which has never been modeled and solved for so far but can be estimated to $\approx$ 7%. It is thus possible to get a good estimate of the solar quadrupole moment: $1.66 \times 10^{-7} \leq J_2 \leq 2.32 \times 10^{-7}$.

**This article has been written in 2017.** Most of the content is still relevant, especially for the Lense-Thirring effect. See also NB at the end of section 3.

## 1 KEY ROLE OF THE SOLAR QUADRUPOLE MOMENTS

Assuming azimuthal symmetry, the solar gravitational moments, $J_n$ are the coefficients in a development on a Legendre polynomials basis of the outer solar gravitational potential at radial distance $r$ from the solar center (with $r \geq R_\odot$). In case of a body obeying the hydrostatic laws, there is the additional north-south symmetry which, together with azimuthal symmetry, implies that odd harmonics are zero ($\forall n, J_{2n+1} = 0$).

The solar quadrupole moments, at first $J_2$ –sometimes called "oblateness" which is an abuse of language and is not physically relevant–, are at the crossroad of solar physics, astrometry and celestial mechanics:

- from the point of view of solar physics, its value reflects the physics of solar models: non-rigid rotation, latitudinal rotation, solar core properties, solar cycle variations (implying a temporal solar dependence of $J_n$), stellar evolution. The solar multipole moments quantify deviations of a gravitational potential from that due a spherical body; they are related to solar shape asphericities

˟ E-mail: jp.rozelot@orange.fr

through the definition of the solar surface as an equipotential surface, with respect to the total potential. Solar shape asphericities, $c_n$ [1] are the coefficients in a development on a Legendre polynomials basis of the shape of internal solar layers at a given depth ($r \leq R_\odot$) (Rozelot et al. 2004, Lefebvre et al., 2007). The $c_{2,4}$ clearly identify the position of characteristic solar internal layers: the core, the tachocline and the so-called leptocline (Godier and Rozelot, 2001, Rozelot et al., 2016.);

- from the point of view of astrometry, high precision astrometry in the solar neighborhood will require a more precise knowledge of the solar quadrupole moment and spin $(\vec{J})$. Indeed, space–time being curved by the solar body, it induces light deflection measured by an angle $\hat{\alpha}$ (Pireaux, 2002), which can be decomposed as[2]:

$$\hat{\alpha}_{\substack{\text{solar grazing incidence}\\\text{light source at infinity}}} = +2(1+\gamma)\frac{GM_\odot}{R_\odot c^2} \quad \approx 1.75(05) \text{ arcsec}$$

$$+ 2(1+\gamma)J_2\frac{GM_\odot}{R_\odot c^2} \quad \approx 0.4 - 0.3 \text{ arcsec}$$

$$\pm 2(1+\gamma)\frac{G|\vec{J}|}{R_\odot c^3} \quad \approx \pm 0.7 \text{ arcsec}$$

$$+ O(\frac{1}{c^4}),$$

where $G$ is the Newtonian constant, $R_\odot$ the solar radius, $M_\odot$ the mass of the Sun, $c$ the speed of light in vacuum, $\gamma$ the Post-Newtonian Parameter encoding the amount of curvature of space–time per unit rest mass. The above estimate of the light deflection was obtained for the time being, in assuming the value $J_2$ of (2.0± 0.4) $10^{-7}$ according to Pireaux and Rozelot (2003). Conversely and in principle, an accurate measurement of the light deflection in the vicinity of the Sun should allow to determine the quadrupole term.

- finally, from the point of view of celestial mechanics, the gravitational moment due to an oblate star plays a role in the relativistic perihelion precession ($\Delta\omega$) of a planet orbiting around this star, and given by:

$$\Delta\omega = \Delta\omega_{(GR)_0} \times [+\frac{1}{3}(2 + 2\gamma - \beta)$$
$$- \frac{R_{star}^2}{R_{Sch_{star}}a(1-e^2)}J_2(3sin^2 i - 1) + 0(J_n)] \quad (1)$$

where $\Delta\omega_{(GR)_0}$ is the general relativistic precession for a mass monopole, *i.e.*:

$$\frac{6\pi GM_{star}}{a(1-e^2)c^2} \; ; \quad (2)$$

$a$, $e$ and $i$ are the usual orbital parameters (semi-major axis, eccentricity, inclination), $R_{Sch_{star}}$, the Schwarschild radius of the oblate star; $\beta$ the Post-Newtonian Parameter encoding the amount of non-linearity in the superposition law of gravitation; $0(J_n)$ indicates that additional terms can be added ($J_n$ as well as shifts due to the perturbation of other planets). For the planets orbiting around the Sun, the total precession per orbit is thus:

$$\Delta\omega = \frac{2\pi GM_\odot}{a(1-e^2)c^2}(2 + 2\gamma - \beta) - \frac{3\pi R_\odot^2}{a^2(1-e^2)^2}J_2(3sin^2 i - 1) + 0(J_n) \quad (3)$$

For Mercury, the shift $\Delta\omega_{(GR)_0}$ is (42.9794 ± 0.0023)″ cy$^{-1}$ (arcsec/century).

Hence a precise knowledge of $J_2$ will be needed for precise ephemerides. Additionally, there are an indirect influence of $J_2$, such as on planetary spins and on the ecliptic plane. Consequently, a precise value of $J_2$ is also needed for long term solar system models (see for instance Laskar et al, 2004).

The order of magnitude of $J_2$ is known to be $\approx 10^{-7}$; such magnitude amounts to about 0.05 % of the general relativistic perihelion advance for Mercury. Precise estimates of $J_2$, however, strongly depend on the method used: the theory of Figures of the Sun, the stellar equations combined with a differential rotation model (from the core to the surface, and depending on the latitudes), or inversion techniques applied to helioseismology (see for instance Rozelot, 2004 and references therein). Nevertheless, a precise value of $J_2$ is useful to compute dynamical effects like the light deflection in the vicinity of the Sun or planetary motion.

Conversely, a precise dynamical estimate of $J_2$ might be crucial to constrain solar density and rotation models. In the following, we consider whether it is possible to extract an estimate of the solar quadrupole moment via planetary ephemerides. In the past, dynamical consequences of $J_2$ already allowed to set constraints on that parameter. Indeed, Moon–Earth spin orbit coupling propagates the influence of the solar $J_2$ to the Moon. Accurate observations of lunar librations lead to the constraint $J_2$ to be less than 3×10$^{-6}$ (Rozelot, 1997; Bois and Girard, 1999).

## 2 EPHEMERIDES TESTS TO DECORRELATE $J_2$ FROM POST NEWTONIAN PARAMETERS

The Parameterized Post–Newtonian PPN formalism dates back to Eddington in 1922, but was fully developed by Nordtvedt and Will (1971) during the years 1968–72. The framework of the theory contains ten PPN parameters. Their precise determinations through accurate observations is one of the major challenge of these latest years to test alternative theories of GR. Indeed the Sun was considered in that way as an experimental laboratory by Brans and Dicke as soon as 1961 (Brans and Dicke, 1961, and a discussion in Damiani et al., 2001; see also a review in Will, 2006 and Ni, 2016). Inspection of Eq. 3 shows that the PN parameters are linked to the quadrupole moment $J_2$ through a linear relation. But the first coefficient varies as 1/$a$ while the second is proportional to 1/$a^2$. The question then becomes to check of what level these two coefficients are correlated, which is not trivial. In the following, we focused on the main PN parameters $\beta$ and $\gamma$, characterizing fully conservative alternative theories of gravitation and the solar quadrupole moment $J_2$.

We present a series of tests to check the correlation between $J_2$ and the PN parameters. The initial values taken for $\beta$, $\gamma$ and $J_2$ are respectively 1.00000, 1.00000, and 2.3 10$^{-7}$ (value allowing a variation range of ± 0.3 10$^{-7}$). Those values of $\beta$ and $\gamma$ correspond to the parametrization of General Relativity, which is still well within the errors bars of Solar system experiments which test gravitational theories. The observed present best constraint on $\gamma$ is

---

[1] Usually, $c_n$ are noted $\gamma_n$, as defined by helioseismology techniques, *i.e.* $\gamma_n$, are the even coefficients of the f-modes. See Rozelot, J.P., Kosovichev, A. and Lefebvre, S., in Highlights of Astronomy, Vol. 14, XXVIth IAU General Assembly, August 2006, K.A. van der Hucht, ed. We emphasize here the need to avoid confusion with the light deflection parameter $\gamma$.

[2] The first term is $\hat{\alpha} = 2\frac{GM}{c^2}(1+\gamma) * \frac{1}{b}$, where $b$ is the impact parameter. The solar case gives, $\hat{\alpha} \approx 8.4*10^{-6}\frac{(1+\gamma)}{2} * \frac{R_\odot}{b}$ (in $rd$).

still given by the Cassini tracking (Bertotti, 2003)[3]:
$\gamma - 1 = (-2.1 \pm 2.3) \, 10^{-7}$ ;
while that on $\beta$ is based on the Lunar Laser Ranging (LLR) technique[4], that is to say (Williams *et al.*, 2009, Equation 27):
$(\beta - 1) \times 10^4 = 1.2 \pm 1.1$
which corresponds to a violation of parameter $\eta = 4\beta - \gamma - 3$.

Solutions to the planetary motion were fitted to observational data by means of the least squares process, in one single step, in order to assign estimates to all unconstrained ephemeris parameters, starting at epoch Julian day 2440400.5. The following observational data of the planets were used: Mercury radar range (1996–70, 1971–82, 1986–97, 1989–97), Venus radar range (1964–65, 1966–70, 1971–95), Venus spacecraft Magellan (VLBI, 1990–94), Mars radar range (1971, 1980, 1982–95), Mars spacecraft Mariner 9 (1971–72), Mars spacecraft Viking (1976–80 and 1980–82), Mars spacecraft Phobos (VLBI, 1989), Mars spacecraft Pathfinder (1997), Jupiter transits (1911-82), Jupiter spacecraft Galileo (VLBI, 1996-97), Saturn transits (1911-82), Uranus transits (1911-1982), Neptune transits (1911-82) and Pluto transits (1989–96). Data are extracted from the Jet Propulsion Laboratory (JPL) Planetary Ephemerides of the Sun, Moon and Planets from 1998 to 2005. The first series of tests (0 to 7 in Table 1) were made in the setting of General Relativity, that is ($\gamma$-1) and ($\beta$-1) constrained to 0. To study their relevance in $J_2$ estimates, Mercury, Venus and Mars planetary data sets were successively taken into account or removed; other data sets were kept. In a second series of tests (8 to 10 in Table 1), all the data sets were taken, but GR was relaxed, and the correlations were studied. Examining Table 1, comments are the followings:
- the resulting estimates of the solar quadrupole moment are mostly sensitive to Mars, Mercury and Venus data, probably through relativistic planetary perihelion advances which call for the contribution of PN parameters $\beta$ and $\gamma$ along with $J_2$;
- estimating $J_2$ in the setting of GR, leads to $(2.2 \pm 0.4) \times 10^{-7}$;
- the comparison between test 0 to 3 shows that when estimating $J_2$
- through planetary ephemerides, Mars ranging is crucial, more than Mercury. As expected, Mercury data are significant as the innermost planet probes the solar gravitational field; considering all data sets and relaxing the general relativistic assumption increases $J_2$. In the setting of fully conservative theories of gravitation (test 10), it is found $J_2 = (2.0 \pm 0.5) \times 10^{-7}$;
- tests 8, 9 and mainly 10, show strong [$\beta$, $J_2$] and [$\gamma$, $J_2$] correlations: $\approx$ 45–78 % and 56–82 % respectively.

The conclusion is that there is a strong correlation between $\beta$, $\gamma$ and $J_2$ in planetary ephemerides, and hence, it is rather difficult to fit these two parameters simultaneously. However, an improved knowledge of the estimate of $J_2$ is crucial in such theories. To disentangle experimentally the solar quadrupole mass moment and the post Newtonian effects, an approach was proposed by Ioro as soon as in 2005 (Iorio, 2005), but was not put into operation to our knowledge. We are still waiting results from space missions for which precision at a level of $10^{-8}$ is expected. We will see in the next section that the ephemeris of planets has been largely improved by adding in the models more and more trajectories of asteroids and trans–neptunian objects; the value of $J_2$ can be thus fitted, but is the result of better accuracy? We believe that an analytic solution as these developed by Xu et al. (2011) to compute the perihelion of planets and compared to the observations could lead to a good alternative for a better estimate of $J_2$.

## 3 PROGRESS MADE THROUGH THE ADVENT OF NEW EPHEMERIDES

High precision of the latest version of planetary ephemerides have been made independently by various authors. The Ephemerides of Planets and the Moon (EPM) were first created in the 1970's in support of Russian space flight missions and were constantly improved at the Institute of Applied Astronomy (IAA–RAS) in Saint-Petersburg (Russia). Pitjeva (2015) reviewed the evolution of the numerical EPM ephemerides from previous released EPM2004, EPM2008, EPM2011, to the new EPM2014 version. The comparison progress of ephemerides includes the growing database of different types of observations from classical optical to radio technical of spacecraft from 1913 to 2014, enlarged up to more 800000 measurements; improved dynamical model from mutual perturbations of all planets, the Sun, the Moon, 301 largest asteroids to additional perturbations of 30 largest trans–neptunian objects. The Jet Propulsion Laboratory (JPL, Pasadena, Ca, USA) Development Ephemerides (DE) remained since a long time the basis of the Astronomical Almanac up to the present. Folkner et al. (2015) develop the Planetary and Lunar Ephemeris DE430 which succeeds the ephemeris DE421 and its precursor DE405. Specific improvements were made including the perturbations from 343 asteroids in the model and a damping term between the Moon's liquid core and solid mantle that gives the best fit to lunar laser ranging data. The "Institut de Mécanique Céleste et de Calcul des Éphémérides" (IMCCE-Paris, F) developed since the early 80's, analytical solutions for the planetary motion called Intégration Numérique Planétaire de l'Observatoire de Paris (INPOP); the latest version available is INPOP15a, an upgraded version of INPOP13c, adding supplementary range tracking data obtained from the analysis of the MESSENGER spacecraft from 2011 to 2014 and including new analysis of Cassini tracking data obtained from 2004 to 2014. The differences obtained by these different teams are regularly confronted between them. The solar quadrupole moment values deduced from the above different procedures can be summarized as follows:

EPM2011 (IAS, Russia): $J_2 = (2.0 \pm 0.2) \times 10^{-7}$ (Pitjeva, 2014)
EPM2013 (IAS, Russia): $J_2 = (2.22 \pm 0.23) \times 10^{-7}$ (Pitjeva, 2014)
INPOP08 (IMCCE, 2008): $J_2 = (1.82 \pm 0.47) \times 10^{-7}$ (Fienga, 2011)
INPOP10a (IMCCE, 2010): $J_2 = (2.40 \pm 0.25) \times 10^{-7}$ (Fienga, 2011)
INPOP10e (IMCCE, 2010) $J_2 = (1.8 \pm 0.25) \times 10^{-7}$ (Verna et al., 2014)
INPOP13a (IMCCE, 2013) $J_2 = (2.40 \pm 0.20) \times 10^{-7}$ (Verna et al., 2014)
INPOP13c (IMCCE, 2013) $J_2 = ((2.11 \text{ to } 2.36) \pm 0.25) \times 10^{-7}$
DE405 (JPL, created May 1997): $J_2 = (1.9 \pm 0.3) \times 10^{-7}$ (Standish, 1998)
DE423 (JPL, created February 2010): $J_2 = (1.8) \times 10^{-7}$ (cited by Verna et al., 2014)
DE430 (JPL, created April 2013): $J_2 = (2.1 \pm 0.7) \times 10^{-7}$ (Folkner et al., 2014)

---

[3] Very Long Baseline Interferometry (VLBI) led to $\gamma - 1 = (1.7 \pm 4.5) \times 10^{-4}$ (Shapiro S., Davis, D., Lebach, J. and Gregory, J.: 2004, Phys. Rev. Lett., 92, 121101).

[4] LLR is a technique in which short pulses of laser light are sent from the Earth to the Moon, reflecting off of arrays of corner cube prisms placed on the Moon's surface by astronauts or unnamed missions. The round-trip time is accurately measured, from which the Earth-Moon distance can be deduced. Comparison with a sophisticated model containing not only gravitational dynamics of the solar system, but also torque effects, Earth tides, surface loading effects, atmospheric propagation delay, etc allows one to test whether GR can adequately describe the lunar orbit, and parametrize any necessary correction (Murphy, 2009).

EPM2004 (IAS, Russia): $J_2 = (1.9 \pm 0.3) \times 10^{-7}$ (Pitjeva, 2005, 2013)
(but $(1.9 \pm 0.6) \times 10^{-7}$ in Pitjeva, 2014)
EPM2008 (IAS, Russia): $J_2 = (1.92 \pm 0.3) \times 10^{-7}$ (Pitjeva, 2014)
EPM2011 (IAS, Russia): $J_2 = (2.0 \pm 0.2) \times 10^{-7}$ (Pitjeva, 2014)
EPM2013 (IAS, Russia): $J_2 = (2.22 \pm 0.23) \times 10^{-7}$ (Pitjeva, 2014)
INPOP08 (IMCCE, 2008): $J_2 = (1.82 \pm 0.47) \times 10^{-7}$ (Fienga, 2011)
INPOP10a (IMCCE, 2010): $J_2 = (2.40 \pm 0.25) \times 10^{-7}$ (Fienga, 2011)

INPOP10e (IMCCE, 2010) $J_2 = (1.8 \pm 0.25) \times 10^{-7}$ (Verna et al., 2014)
INPOP13a (IMCCE, 2013) $J_2 = (2.40 \pm 0.20) \times 10^{-7}$ (Verna et al., 2014)
INPOP13c (IMCCE, 2013) $J_2 = ((2.11 \text{ to } 2.36) \pm 0.25) \times 10^{-7}$
First range of values: if $\beta$ and $\gamma$ are fixed to one and $\dot{G}/G$ to be equal to zero; second value: in an adjustment with new B2 recommendation of the IAU–GA in 2015.

| Test | Data | $J_2 \times 10^{-7}$ | $(\beta-1) \times 10^{-4}$ | $(\gamma-1) \times 10^{-4}$ | $\dfrac{J_2}{\beta}$ | $\dfrac{J_2}{\gamma}$ | $\dfrac{\beta}{\gamma}$ |
|---|---|---|---|---|---|---|---|
| 0 | All data (nominal GR solution) | 2.2 ± 0.4 | 0 | 0 | | | |
| 1 | No Mercury data | 3.6 ± 0.8 | 0 | 0 | | | |
| 2 | No Venus data | 2.2 ± 0.4 | 0 | 0 | | | |
| 3 | No Mars data | 3.1 ± 1.2 | 0 | 0 | | | |
| 4 | No Mars ranging data | 0.5 ± 0.6 | 0 | 0 | | | |
| 5 | No Mars VLBI data | 2.6 ± 0.4 | 0 | 0 | | | |
| 6 | No early Mars data (pre–1985) | 2.3 ± 0.4 | 0 | 0 | | | |
| 7 | No late Mars data (post–1985) | 2.0 ± 0.5 | 0 | 0 | | | |
| 8 | All data | 3.2 ± 0.7 | 0 | -2.5 ± 2.3 | | - 0.82 | |
| 9 | All data | 2.7 ± 0.6 | 1.0 ± 3.2 | 0 | + 0.78 | | |
| 10 | All data (fully conservative) | 2.0 ± 0.5 | -0.1 ± 3.4 | -2.6 ± 2.4 | + 0.45 | - 0.56 | + 0.30 |

**Table 1.** *Ephemerides tests, starting epoch Julian day 2440400.5, carried out by removing or including planets motion. The first column gives the number of case tests, as defined in column two. Third to fifth column provide the solution values for $J_2$ and Post–Newtonian Parameters $\beta$, $\gamma$. Sixth to eighth columns give the correlation between pairs of these three quantities. The zero inside a column means that the value was constrained. These tests have been performed to check the dependence of $J_2$ to the influence of planets (in the frame of GR) and to evaluate the strong correlation between the pairs $[\beta, J_2]$ and $[\gamma, J_2]$.*

Such values can be compared with our estimate deduced from the best observations available at that time, $J_2 = (2.00 \pm 0.40) \times 10^{-7}$ (Pireaux and Rozelot, 2003). A table of available values is given in Pitjeva (2013) and Damiani and Rozelot (2011). If we consider that the advances in techniques for deep space exploration in the solar system, have increasingly improved ephemerides with high–precision datasets provided from tracking spacecrafts and by sophisticated data analysis, the ponderated value (according to
$v = \sum(v_i/\sigma_i^2)/\sum(1/\sigma_i^2)$; $\sigma = (1/\sum(1/\sigma_i^2))^{0.5}$ for experimental values v of error $\sigma$; see Lyons, L. Data analysis, Cambridge University Press, 1992, p. 31) deduced from DE430 and EPM2013 (the most recent ones; albeit all data, except DE423, leads to $J_2 = (2.10 \pm 0.08) \times 10^{-7}$) is:

$$J_2 = (2.21 \pm 0.22) \times 10^{-7}.$$

**NB: New data are available in Rozelot and Eren, 2020, Sp. Sc. Res.,65, 2821.**

## 4 CONSTRAINT DUE TO THE LENSE-THIRRING PRECESSION

Predicted only a few years after the formulation of General Relativity, the Lense-Thirring effect (Lense and Thirring, 1918) demonstrates the embodiment of Mach's principle in this theory (Rindler, 1994). As a result of this effect, the lines of nodes of a particle orbiting a spinning star is expected to drift in the direction of star rotation, leading to a secular precession (prograde) of the longitude of the ascending node (Stella, 2009). The Lense–Thirring precession rate is slow around the Sun and extremely slow around the Earth. Such non–modeled gravitomagnetic effect are partially or totally removed from the post–fit signature in the data–reduction process of the planetary and lunar ephemerides as previously seen. According to Iorio (2005) the Lense–Thirring effect consists of tiny secular precessions of the longitude of the ascending node $\dot{\Omega}$ and the argument of pericentre $\dot{\omega}$ of the orbit of the test particle:

$$\dot{\Omega}_{LT} = \frac{2GJ_\odot}{c^2 a^3 (1-e^2)^{3/2}} \qquad \dot{\omega}_{LT} = \frac{6GJ_\odot \cos i}{c^2 a^3 (1-e^2)^{3/2}} \qquad (4)$$

where $a$, $e$ and $i$ are respectively the semimajor axis, the eccentricity and the inclination of the orbit, $c$ is the speed of light, $G$ is the Newtonian gravitational constant and $J_\odot$ the Sun's angular momentum which can be taken as $1.9 \times 10^{41}$ kg m$^2$ s$^{-1}$. Eq. ?? holds only in a coordinate system whose $z$ axis is aligned with the Sun's angular momentum. Iorio (2012-a and b) analytically worked out the secular Lense–Thirring precessions of the Keplerian orbital elements for a generic orientation of the angular momentum of the central body, in a form that allows easier comparisons with the observation–based quantities usually determined by the astronomers. Taking into account the Lense–Thirring effect, for which numerical computations lead to an estimate of $\approx$ 7%, we can infer a value of the solar quadrupole moment $J_2$ as (in $10^{-7}$) 2.06 ± 0.6 according to EPM2013 and 1.96 ± 0.4 according to DE 430, so that the ponderated value can be set up as:

$$J_2 = (1.99 \pm 0.33) \times 10^{-7}.$$

The resulting limits on the solar gravitational moment is thus particularly constraint.

## 5 CONCLUSION

It has been shown since a long time that a gravitational quadrupole moment of the Sun influences the motion and inclination of the planetary orbits. Up to the beginning of the year 2014, the range of $J_2$ was set up from $1.4 \times 10^{-7}$ to $2.4 \times 10^{-7}$. Due to the progress in ephemerides computations, today, one can restrict $J_2$ to $1.99 \times 10^{-7}$ with an error of $\pm 0.33 \times 10^{-7}$. With new ground–based observations using radar astrometry (at Arecibo by Margot et al. 2009) in the inner solar system, scheduled to the next coming years, it is expected to provide a purely dynamical measurement of $J_2$ at the $10^{-8}$ level. The same conclusion arises with the incoming new spacecrafts data, which will provide an unprecedented number of observations of perihelion shifts. However, the temporal variation of this parameter is not definitely known either, even if constraints can be put.

NB: The improvement will come from Bepi-Colombo and maybe also from Gaia.

## 6 ACKNOWLEDGMENTS

The authors thank Sophie Pireaux for her computations and meaningful interactions that have contributed to the comprehension of the physics hidden in solar gravitational moments. One of us (JPR) also acknowledges the International Space Science Institute in Bern (Switzerland) for a "visitor scientist" grant. Much of the present paper has been written and discussed there.